# SOXS End-to-End simulator: development and applications for pipeline design.

M. Genoni*[a], M. Landoni[a,u], G. Li Causi[b], G. Pariani[a], M. Aliverti[a], S. Campana[a], P. Schipani[c], R. Claudi[d], M. Munari[k], A. Rubin[o], P. D'Avanzo[a], M. Riva[a], A. Baruffolo[d], F. Biondi[t], G. Capasso[c], R. Cosentino[f], F. D'Alessio[g], O. Hershko[e], H. Kuncarayakti[h,i], G. Pignata[l,m], S. Scuderi[k], K. Radhakrishnan[d], S. Ben-Ami[e], F. Vitali[g], D. Young[o], A. Brucalassi, J. Achrén[p], J. A. Araiza-Duran[q,m], I. Arcavi[r], R. Bruch[e], E. Cappellaro[d], M. Colapietro[c], M. Della Valle[c], M. De Pascale[d], R. Di Benedetto[k], S. D'Orsi[c], A. Gal-Yam[e], M. Hernandez Diaz[f], J. Kotilainen[h,i], S. Mattila[i], M. Rappaport[e], D. Ricci[d], B. Salasnich[d], S. Smartt[o], R. Zanmar Sanchez[k], M. Stritzinger[s], H. Ventura[f]

[a] INAF– Osservatorio Astronomico di Brera-Merate, via E. Bianchi 46, I-23807 Merate (LC), Italy;
[b] INAF– Istituto di Astrofisica e Planetologia Spaziali, via Fosso del Cavaliere 100, Roma – Italy;
[c] INAF – Osservatorio Astronomico di Capodimonte, Sal. Moiariello 16, I-80131, Naples, Italy
[d] INAF – Osservatorio Astronomico di Padova, Vicolo dell'Osservatorio 5, I-35122, Padua, Italy
[e] Weizmann Institute of Science, Herzl St 234, Rehovot, 7610001, Israel;
[f] FGG-INAF, TNG, Rambla J.A. Fernández Pérez 7, E-38712 Breña Baja (TF), Spain;
[g] INAF – Osservatorio Astronomico di Roma, Via Frascati 33, I-00078 M. Porzio Catone, Italy;
[h] Finnish Centre for Astronomy with ESO (FINCA), FI-20014 University of Turku, Finland;
[i] Tuorla Observatory, Dept. of Physics and Astronomy, FI-20014 University of Turku, Finland;
[k] INAF – Osservatorio Astrofisico di Catania, Via S. Sofia 78 30, I-95123 Catania, Italy;
[l] Universidad Andres Bello, Avda. Republica 252, Santiago, Chile;
[m] Millennium Institute of Astrophysics (MAS), Santiago, Chile;
[n] ESO, Karl Schwarzschild Strasse 2, D-85748, Garching bei München, Germany;
[o] Astrophysics Research Centre, Queen's University Belfast, Belfast, BT7 1NN, UK;
[p] Incident Angle Oy, Capsiankatu 4 A 29, FI-20320 Turku, Finland;
[q] Centro de Investigaciones en Optica A. C., 37150 León, Mexico;
[r] Tel Aviv University, Department of Astrophysics, 69978 Tel Aviv, Israel;
[s] Aarhus University, Ny Munkegade 120, D-8000 Aarhus, Denmark;
[t] Max-Planck-Institut für Extraterrestrische Physik, Giessenbachstr. 1, D-85748 Garching, Germany;
[u] INAF– Osservatorio Astronomico di Cagliari, via della Scienza 5, 09047, Selargius (CA), Italy;

**ABSTRACT**

We present the development of the End-to-End simulator for the SOXS instrument at the ESO-NTT 3.5-m telescope. SOXS will be a spectroscopic facility, made by two arms high efficiency spectrographs, able to cover the spectral range 350-2000 nm with resolving power R≈4500. The E2E model allows to simulate the propagation of photons starting from the scientific target of interest up to the detectors. The outputs of the simulator are synthetic frames, which will be mainly exploited for optimizing the pipeline development and possibly assisting for proper alignment and integration phases in laboratory and at the telescope. In this paper, we will detail the architecture of the simulator and the computational model, which are strongly characterized by modularity and flexibility. Synthetic spectral formats, related to different seeing and observing conditions, and calibration frames to be ingested by the pipeline are also presented.

*send correspondence to: matteo.genoni@inaf.it;

# 1. INTRODUCTION

End-to-End instrument models (E2E) are numerical simulators, which aim to simulate the expected astronomical observations starting from the radiation of the scientific sources (or calibration sources in the case of calibration frames) to the raw-frame data produced by instruments. Synthetic raw-frames can be ingested by the Data Reduction Software (DRS) to be analyzed in order to assess if top-level scientific requirements, such as spectral resolution, SNR, Radial Velocity precision, related to the specific science drivers, are satisfied with the specific instrument design and architecture.

E2Es have been valuable software exploited in many types of astronomical instruments for different purposes. Specifically, they have been used in design phases to optimize and improve specific hardware components and parameters[1], or for early verification of instrument performance[2]. From the scientific point of view, they have been extensively exploited for assessing the feasibility of particularly challenging observations[3]. Furthermore, instrument simulators are systematically exploited to aid both Data Reduction[4] and Data Analysis Software development, as well as for testing and verifying of existing data reduction pipelines. In fact, in the specific case of SOXS, the E2E simulator, which will be here presented, has a key goal to drive the optimization of the DRS development, especially in the UV-VIS arm. In addition, it will also be a reliable tool to help the setting of calibration procedures and observation plans.

SOXS will be a spectroscopic facility, made by two arms high efficiency spectrographs, able to cover the spectral range 350-2000 nm with resolving power R≈4500. It is in construction phase and will be installed at the ESO-NTT 3.5-m telescope (see for a wide overview[6,7]).

This paper is organized as follows. In the Section 2, an overview of the SOXS instrument is given. The Simulator architecture is described in section 3, while a set of simulated frames is discussed in Section 4.

# 2. THE SOXS INSTRUMENT

SOXS is a wide-band spectrograph for the NTT (it will be installed at one of the Nasmyth foci of the NTT) covering in a single exposure the spectral range from the UV to the NlR (350-2000 nm). Its central structure (the backbone) supports two distinct spectrographs, one operating in the UV-VIS 350-850 nm and the other in the NIR 800-2000 nm wavelength ranges. Both spectrographs can operate at different resolutions according to the slit widths: R≈10000 with 0.5" slit, R≈4500 with 1" slit and R≈3300 with 1.5" slit.

The two arms are fed by the light coming through a common opto-mechanical system (the Common Path[8]). It redirects the light from the telescope focus to the spectrograph slits through relay optics reducing the F/number (from F/11 to F/6) and compensating for the atmospheric dispersion (only in the UV-VIS). The Common Path provides also the mechanism to drive the light to/from the other instrument subsystems, i.e. the acquisition camera and the calibration unit.

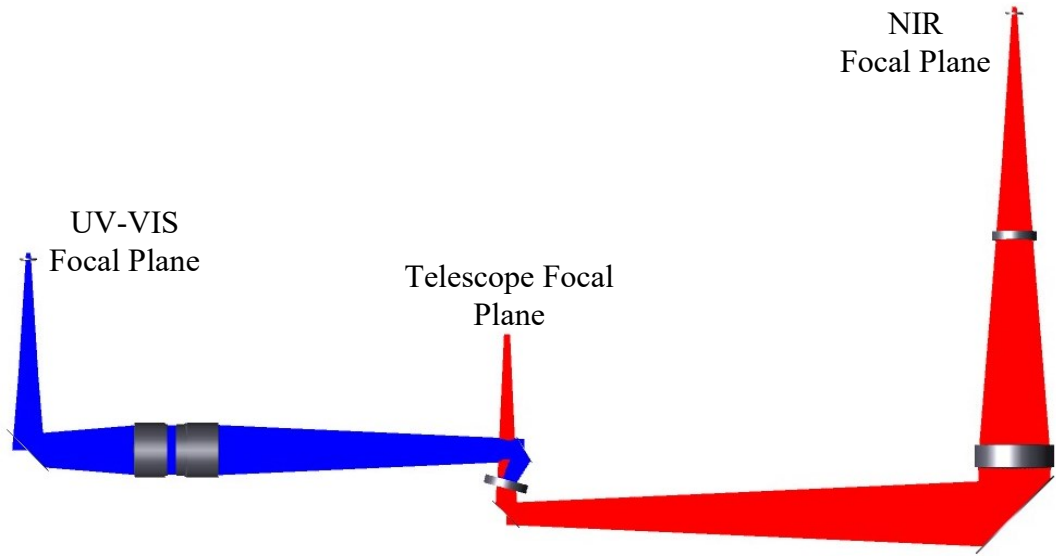

Figure 1. SOXS Common-Path complete optical layout. On the left (in blue) the UV-VIS arm, on the right (in red) the NIR arm.

The UV-VIS spectrograph arm is based on a novel multi-grating concept (for details[9] and Fig. 2), in which the incoming beam is partitioned into four polychromatic beams using dichroic surfaces, each covering a waveband range of ~100 nm, named as quasi-order. Each quasi-order is diffracted by an ion-etched grating. The detector, located 4 mm behind the camera field flattener back surface, is an e2V CCD44-82 CCD (2048 x 4096 pixels, Pixel size 15 μm, see for details[10]).

The near infrared spectrograph, showed in Fig. 3, is a cross-dispersed echelle, with R~5000 (for 1*arcsec* slit), covering the wavelength range from 800 to 2000nm with 15 orders (see for details[11]). It is based on the 4C concept, characterised by a very compact layout, reduced weight of optics and mechanics, good stiffness. The spectrograph is composed of a double pass collimator and a refractive camera, a R-1 grating as main disperser and a prism-based cross disperser. The detector is a Teledyne H2RG array operated at 40K, 2048 x 2048 pixels, Pixel size 18 μm (see for details[12]).

The calibration unit, provides the calibration spectra to remove the instrument signature. The calibration spectra are generated using a synthetic light source, adopting an integrating sphere equipped with lamps suitable for wavelength and flux calibrations across the full wavelength range of the instrument (350-2000 nm); for details see[13]. The following lamps are used:

- Quartz-tungsten-halogen (QTH) lamp, for flat field frames 500-2000 nm;
- Deuterium (D2) lamp, for flat field 350-500 nm (used simultaneously with QTH lamp for UV-VIS arm);
- Ne-Ar-Hg-Xe pen-ray lamps bundled together, for NIR wavelength calibration. The individual lamps are controlled to operate together as one lamp;
- ThAr hollow cathode lamp, for UV-VIS wavelength calibration.

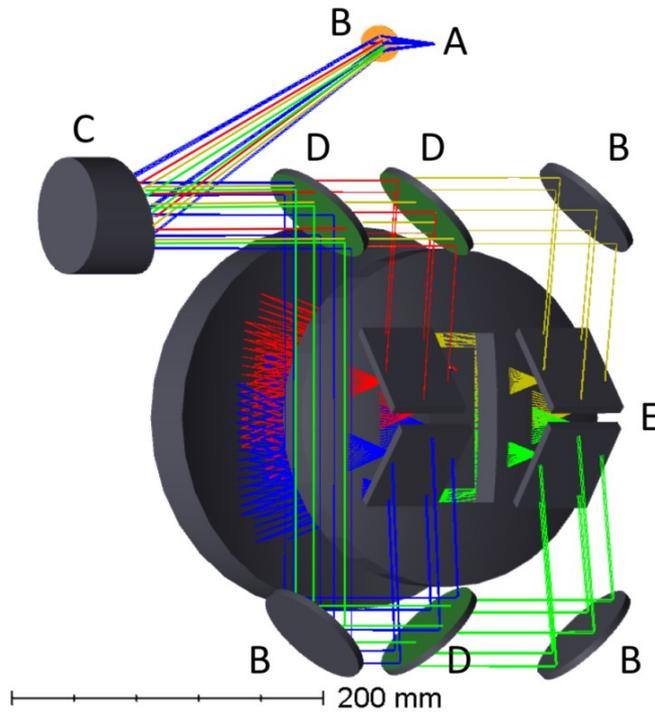

Figure 2. SOXS UV-VIS spectrograph complete optical layout. A. slit plane, B. reflective mirrors, C. OAP Collimator, D. dichroic filters, E. gratings.

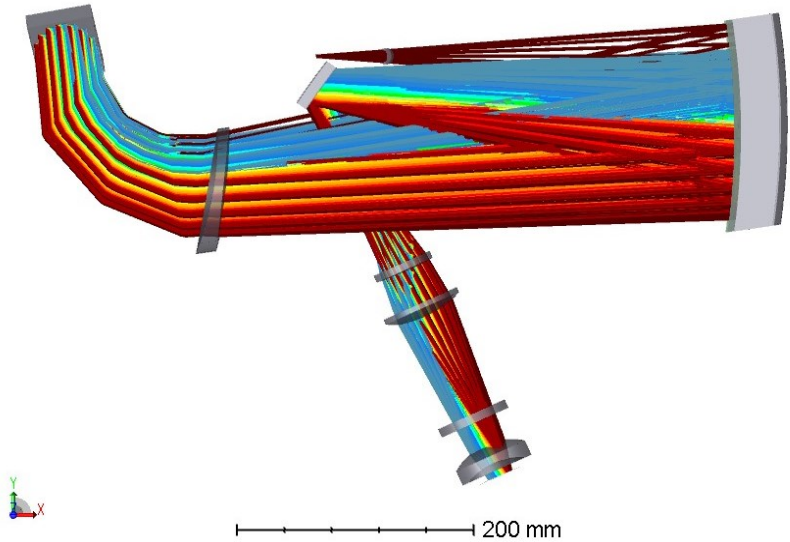

Figure 3. SOXS NIR spectrograph complete optical layout. The optical beams goes the double pass collimating and cross-dispersing optics. The main disperser is a 44deg blaze angle echelle grating. A field mirror redirect the beam down to the camera.

## 3. THE SIMULATOR ARCHITECTURE

The End-to-End simulator architecture is highly modular, composed by different modules (each one with specific tasks and functionalities) units and interfaces, as described in the schematic workflow of Fig. 4. Modularity and flexibility are key points in the definition of tasks and interfaces among different modules or units to allow this tool for being scalable and adaptable to different kind of simulations (science versus calibrations frames) and spectroscopic instrumentation. Modules and Units are characterized by the main tasks for which they are in charge, the required inputs and expected outputs (in a specified format). The simulator is written in MATLAB® and uses specific libraries for specific functionalities and for interfacing with other software, like commercial optical ray tracing Zemax-OpticStudio®.

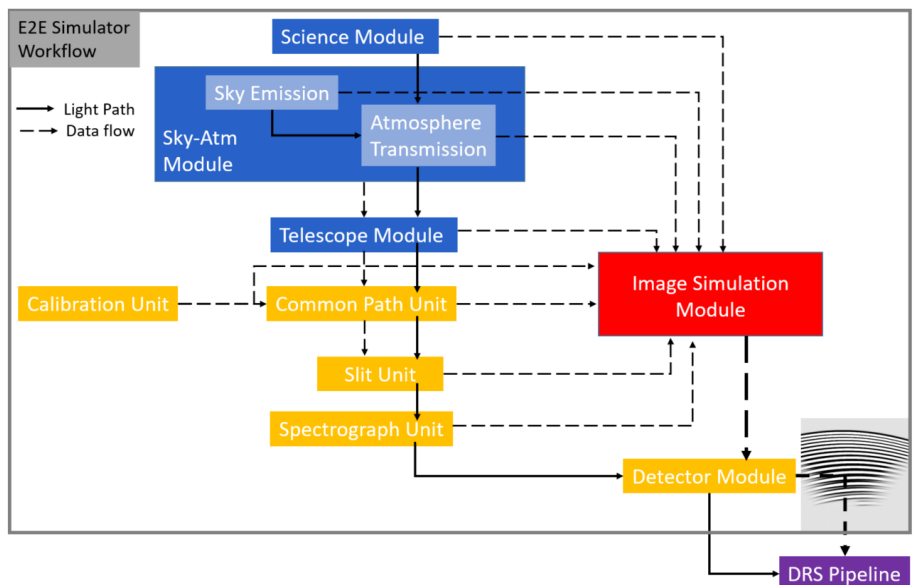

Figure 4. E2E simulator workflow schematic description. The solid arrows represent the real light path, while the dashed arrows show: the simulation data-flow, how the different modules and units are interfaced and their connection to the simulator core, which is the Image Simulation Module. Yellow blocks are units of the Instrument Module, while blue blocks are related to simulation modules independent from the specific instrument.

In the following a brief description of the modules and units showed in the workflow diagram is given.

### 3.1 Science Module

The Science Object Module has the purpose to generate a synthetic 1D spectrum related to a specific astronomical source, at a resolution higher than the one selected for the instrument simulation (which is related to the instrumental resolution dependent on the slit width). The 1D spectral flux density distribution is obtained by a set of parameters (e.g. object type - spectral type for stars -, magnitude and red-shift) or by loading the spectrum from a user-defined library. Once the spectrum is loaded, it is first shifted in wavelength according to the red-shift, then it is normalized at the reference wavelength of the provided magnitude pass-band and then re-scaled accordingly. The output of this module is a FITS table containing the spectral flux density, in units of photon flux, at each wavelength. An example of a synthetic 1D spectrum for a G0V star (Pickels model) sampled at resolution of 0.5Å, is shown in the following figure.

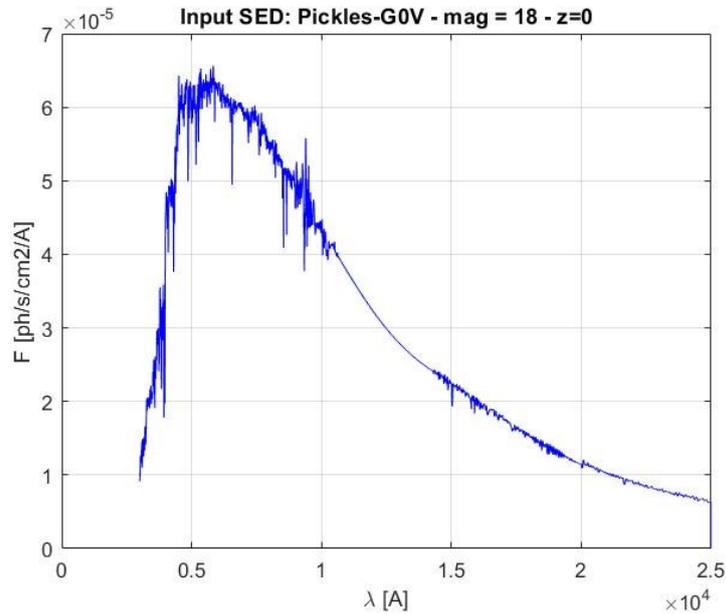

Figure 5. Example of a synthetic 1D spectrum for a G0V star (Pickels model) sampled at resolution of 0.5Å, mag 18 (*V* band of Johnson-Cousins filters, in Vega mag-System) and *z*=0.

### 3.2 Sky-Atmosphere module

The task of this module is to model the scattering, absorption and emission occurring in the Earth's atmosphere. This is done by calling sky-emission and atmospheric transmission spectra of a dedicated library built using the ESO SkyCalc tool (available at the web page[14]), which is based on the Cerro Paranal Advanced Sky Model. These spectra can be loaded from the library according to the sky conditions parameters set for the simulation, i.e. moon phase, air-mass (AM) and precipitable water vapor (PWV). The sky radiance spectrum loaded is in units of ph/s/$m^2$/μm/$arcsec^2$, thus it is first calculated for the on-sky area related to the slit length and selected slit width according to the simulation (both in arcsec) and then the units are converted in ph/(s $cm^2$ Å).

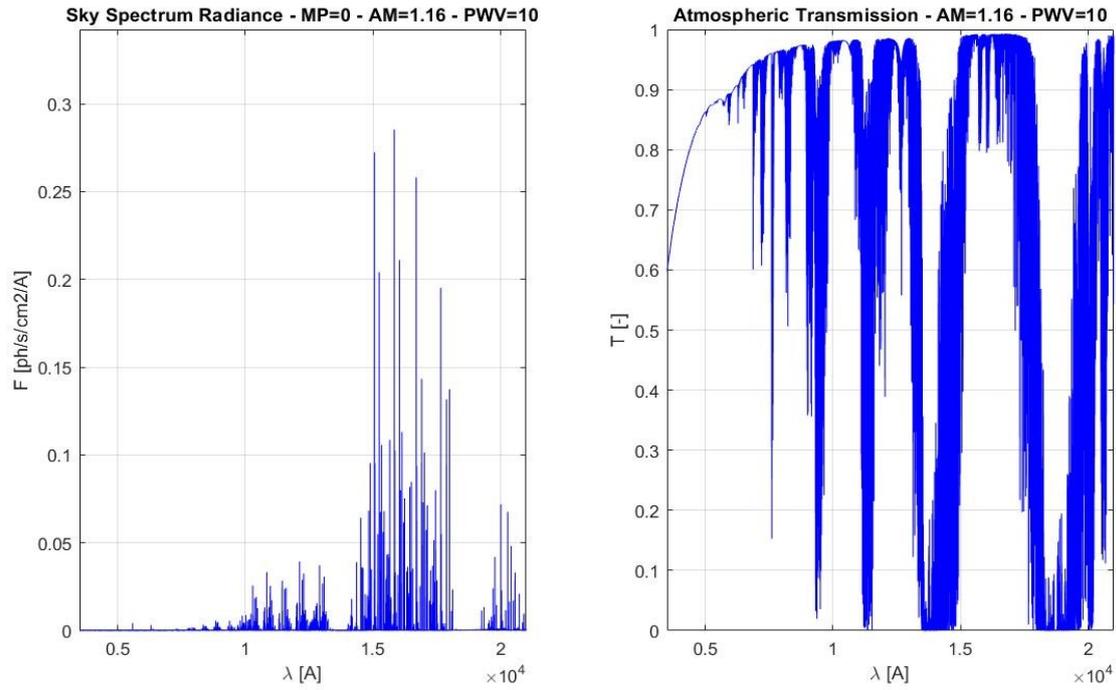

Figure 6. Example of a Sky radiance spectrum and atmospheric transmission calculated through the ESO-SkyCalc tool. Moon phase is 0 (i.e. new moon), airm-mass = 1.16 and precipitable water vapor (PWV) = 10 mm.

### 3.3 Telescope Module

The aim of this module is to predict the telescope throughput based on the available telescope mirrors reflectivity data. For the NTT a global telescope throughput of 0.5 is considered.

## 3.4 Calibration Unit

The calibration unit is in charge to simulate the Spectral Energy Distribution (SED) of the calibration sources. This unit requires to know which kind of lamps are to be simulated flats lamp or ThAr and pen-rays lamps according to the specific calibration frame to be generated, the instrument resolution and the type of calibration mask. In fact, SOXS foresees a single pinhole and a multi pinholes (9 pinholes) mask for calibrations and DRS purposes. In the following examples of the QTH spectrum (in UV-VIS) and PenRays (Ar-Hg-Ne-Xe) spectrum in the NIR.

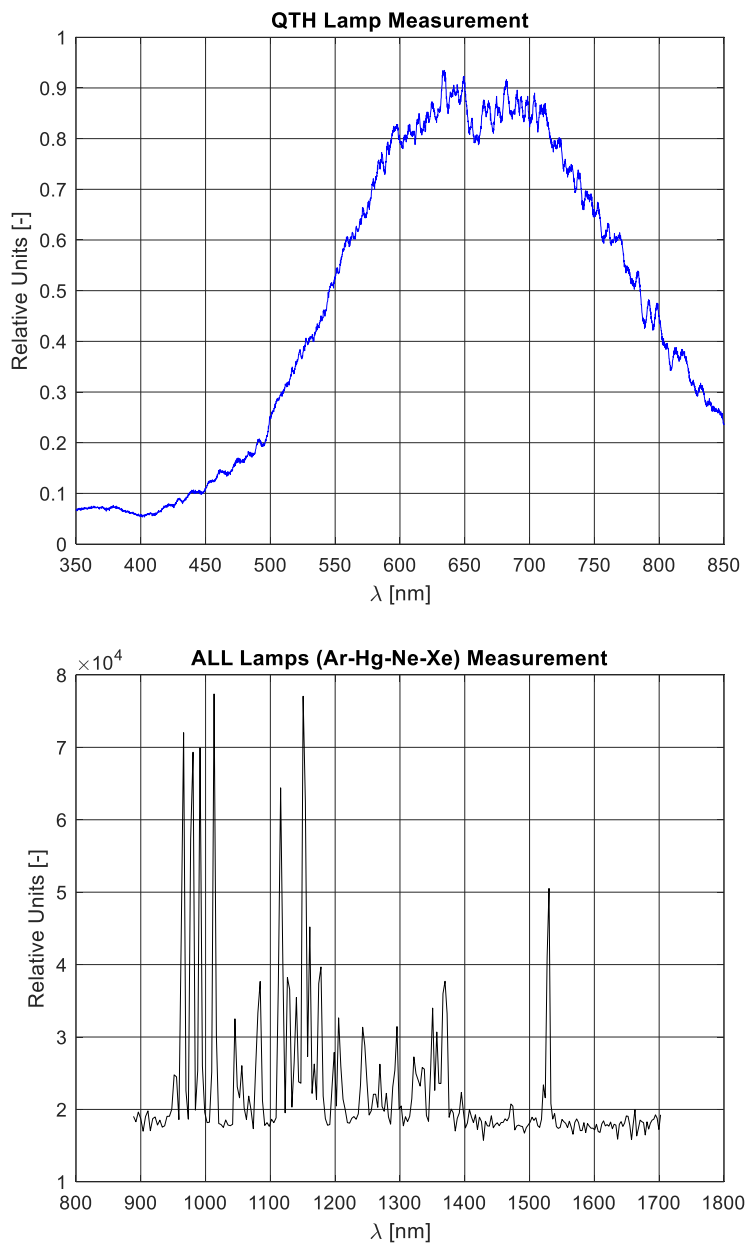

Figure 7. Example of calibration sources spectra used in SOXS. Top panel QTH spectrum, Bottom panel Penrays. Both spectra have been measured by the Calibration Unit work-package team.

## 3.5 Common Path Unit and Slit Unit

The task of the Common-Path (CP) unit is to predict the light distribution at the CP focal plane and throughput for both the UV-VIS arm and NIR arm. The required inputs are seeing values for the observing conditions, the CP image scale and the optics data regarding glasses, coatings and mirrors reflectivity in order to estimate the image on the CP focal plane and the efficiency. The UV-VIS and NIR common path arms efficiency has been calculated combining both measurements data from manufacturer test reports, for the already built optical elements, and estimations for the components that are currently under procurement.

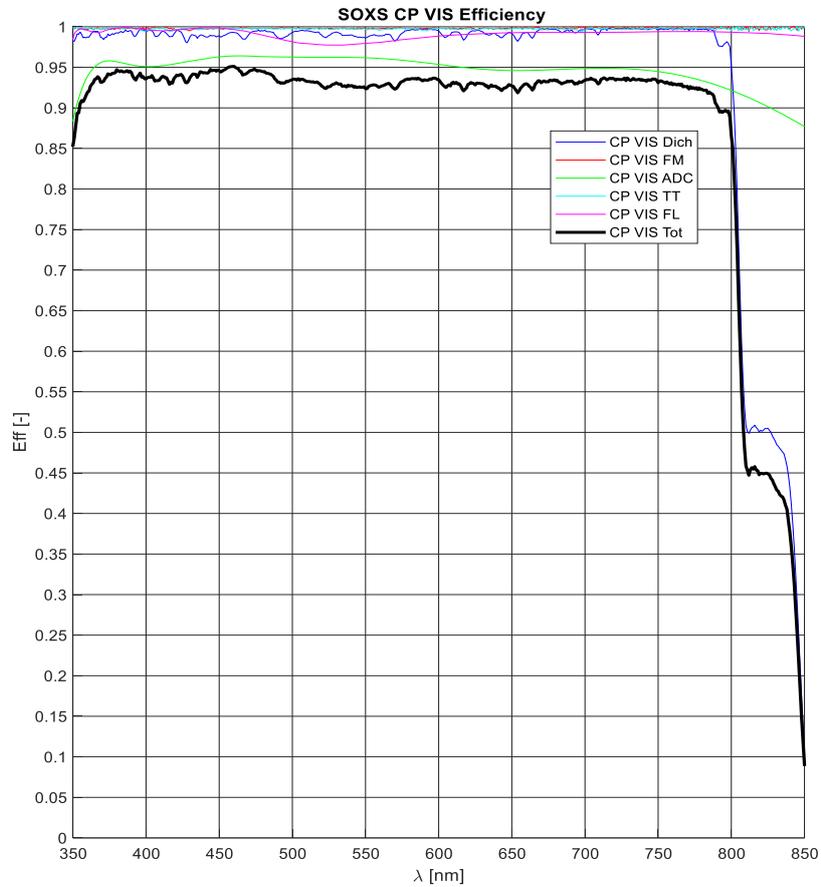

Figure 8. Common-Path UV-VIS arm efficiency, in black bold line, while in colored lines the single Common-Path optical elements efficiency. Dich – Dichroic, FM – folding mirror, ADC – atmospheric dispersion corrector, TT – tip-tilt mirror, FL – field lens.

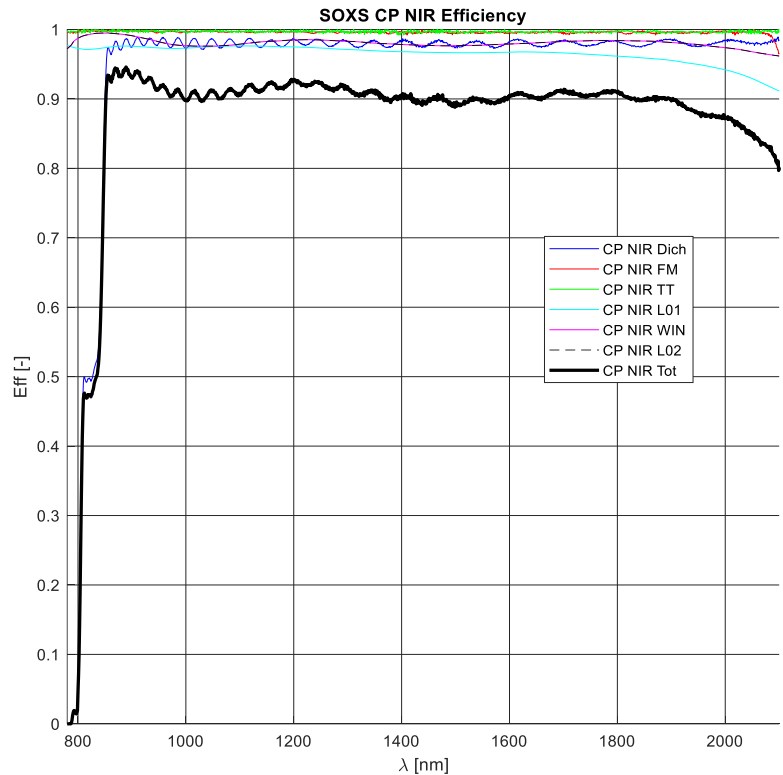

Figure 9. Common-Path NIR arm efficiency, in black bold line, while in colored lines the single Common-Path optical elements efficiency. Dich – Dichroic, FM – folding mirror, TT – tip-tilt mirror, L01 – lens 1 (refocuser), WIN – window, L02 – lens2 (field lens).

The slit-unit calculates the fraction of light passing through the UV-VIS and NIR slits, which are located on their respective common-path arms focal plane, from the variation of seeing in wavelength (w.r.t. the reference value set at 5000 Å) and considering the specific sizes (the slit length is constant 12", while the width changes according to the simulated set of instrument parameters).

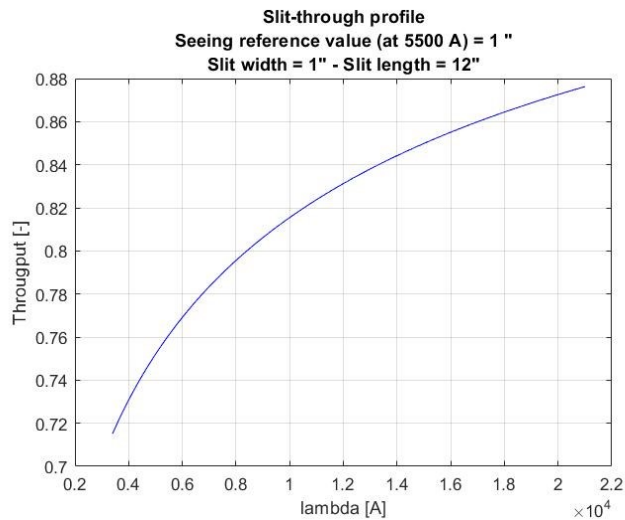

Figure 10. Slit throughput profile for a seeing value of 1" (at reference wavelength of 5500 A) and the 1" slit width (for which the SOXS resolving power is about 4500).

## 3.6 Spectrograph Unit

The purpose of this Unit is to simulate the physical effects of the different optical components of the spectrometers with the final aim of predict the echellogram (spectral format) at the UV-VIS and NIR focal plane, the throughput and a database of PSF maps for a set spectral resolution element (which are used by the Image simulation Module for rendering the synthetic frames). Aberrations, distortion and diffraction effects have already been taken into account at this current simulator version, while the physical operative conditions of the instrument in term of mechanical and thermal effects will be introduced in the future development versions according to the required simulation scenarios.

As for the common path, also the spectrograph arms efficiency is calculated combining both measurements data from manufacturer test reports, for the already built optical elements, and estimations for the components that are currently under procurement. UV-VIS is plotted in Fig. 11, while NIR in Fig. 12.

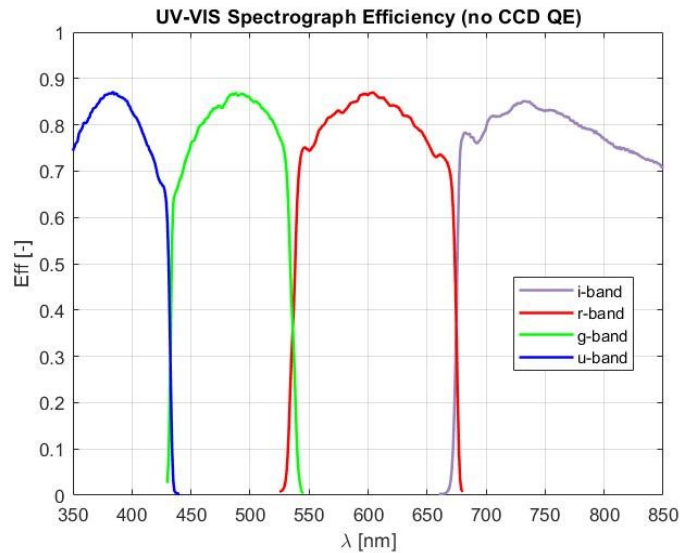

Figure 11. UVVIS spectrograph quasi orders efficiency curves. These are used for the computation of the single quasi-orders synthetic spectrum on the CCD.

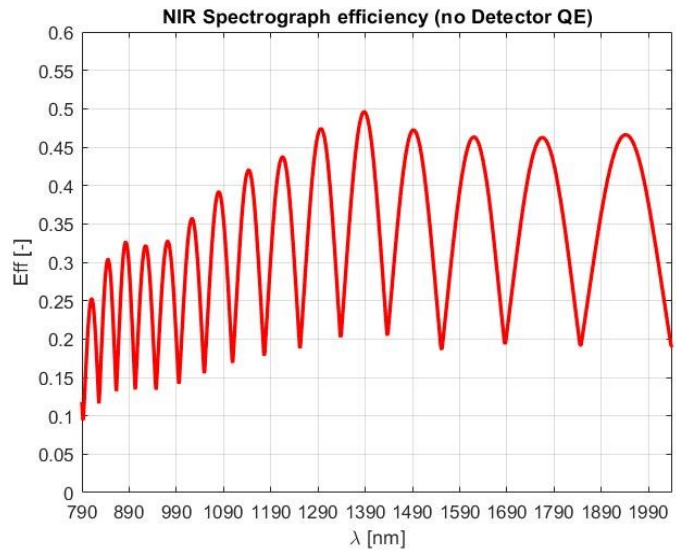

Figure 12. NIR spectrograph global efficiency curve.

The PSF maps database is computed by directly querying the optical design ray-tracing files (built with commercial optical design software Zemax-OpticStudio®), using specific API-based software codes to extract the required data for a defined set of wavelengths along each diffraction order (or the for quasi-orders in the UV-VIS arm). In particular, the PSF maps are calculated using the Huygens PSF analysis tool of Zemax-OpticStudio®, which includes diffraction and optical aberrations. Being SOXS a long slit spectrograph the PSF map for each database wavelength is computed from an average of the PSF maps, 64μm wide, related to five positions along the slit (-0.662, -0.331, 0, 0.331, 0.662 mm, corresponding to -6, -3, 0, 3, 6 *arcsec*). Examples for both arms, first UV-VIS then NIR, in the following figures-tables.

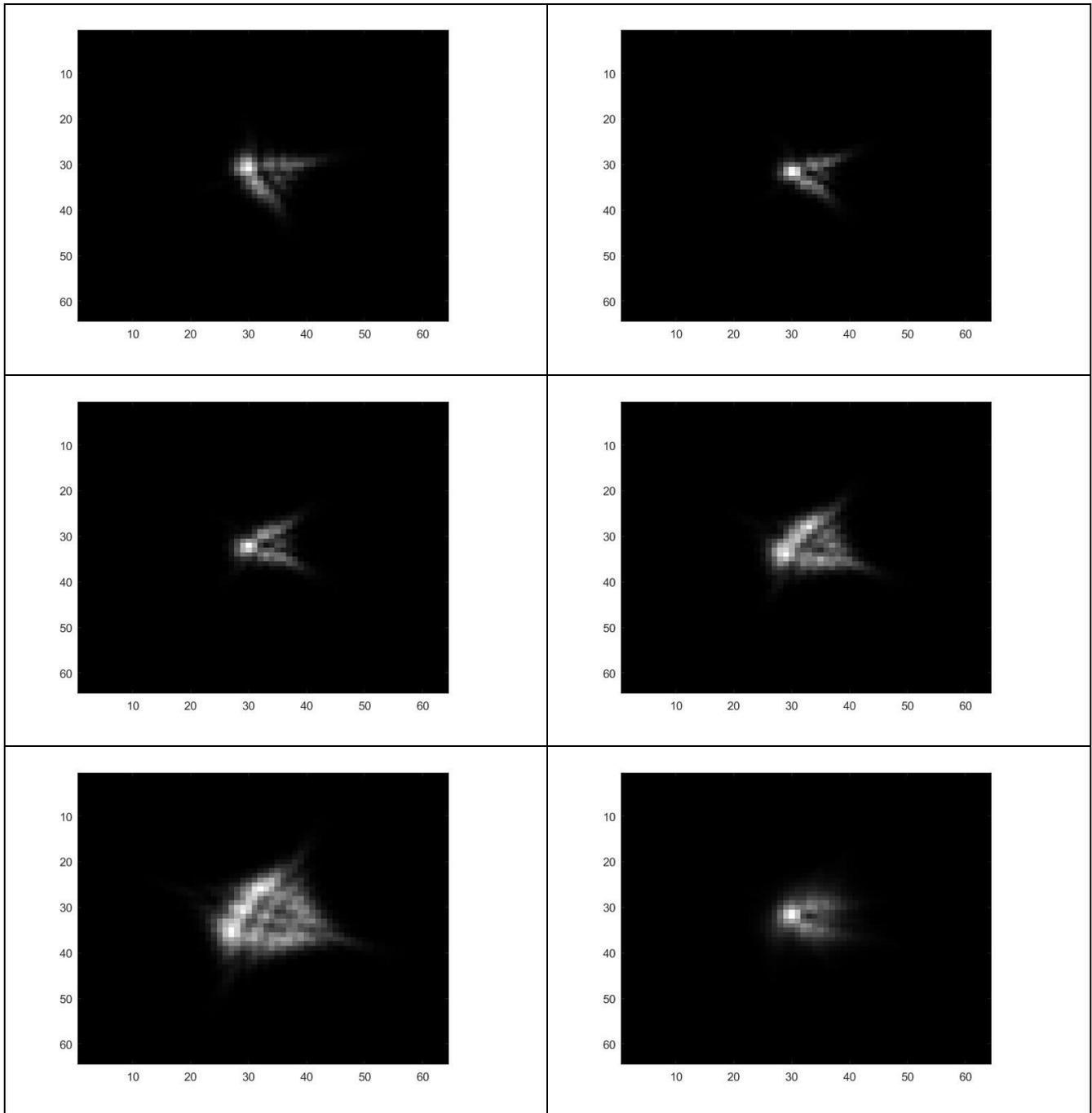

Figure 13. UV-VIS spectrograph example of PSF maps, 64 μm wide, computed for five positions along the slit (-0.662, -0.331, 0, 0.331, 0.662 mm) and the average PSF which is then saved in the database.

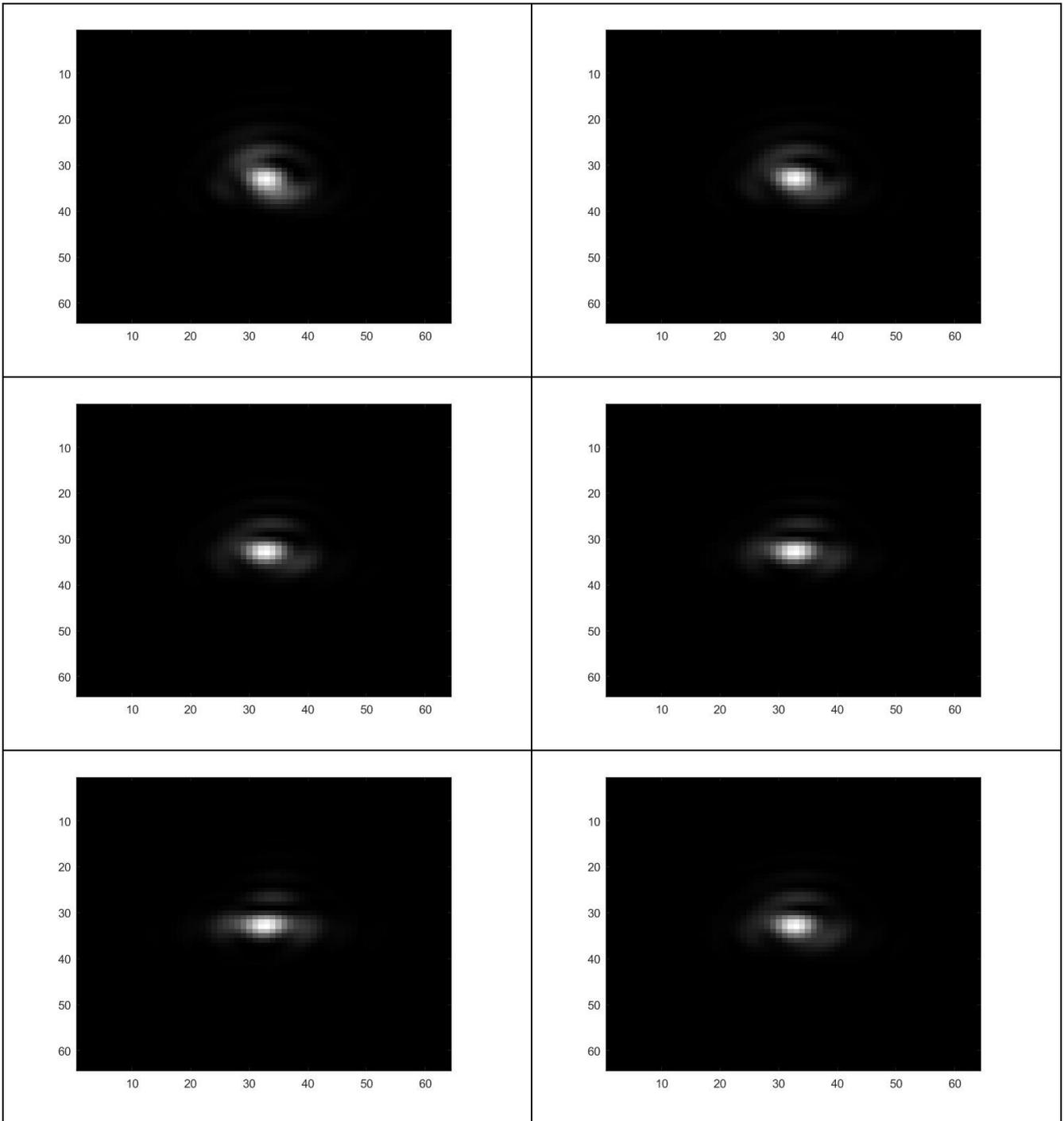

Figure 14. NIR spectrograph example of PSF maps, 64 μm wide, computed for five positions along the slit (-0.662, -0.331, 0, 0.331, 0.662 mm) and the average PSF which is then saved in the database.

The spectral format data, required to produce the synthetic echellograms by the Image Simulation Module, are the diffraction order number (or quasi order id, for UV-VIS), wavelength (the same set used in the computation of the PSF maps database), centroid coordinates of both the central slit position and edges (through which the slit image tilt on the focal plane is computed) and slit image size (i.e. the geometrical full width in main dispersion and spatial direction).

The full database, comprised of efficiency, PSF maps and spectral format, produced by the Spectrograph Unit is stored for the usage of the Image Simulation Module

### 3.7 Image Simulation Module

This portion of the simulator is the kernel of the whole system and put together the outputs of all the other modules and units. This piece of software is responsible for rendering the photons distribution of each spectral resolution element for each order as should be detected at the level of spectrographs focal plane by the detectors. In particular, the module first interpolates on a sub-pixel scale, of which the oversampling can be set according to the required simulation accuracy, all the instrumental data regarding wavelength, image centroid coordinates, sampling (in both main dispersion and spatial directions), slit image tilt, average PSF map and efficiency. Then, it produces the photons distribution of each wavelength in sub-pixel scale by convolving the slit (long slit or pinhole masks according to the specific science or calibration frame type) image with the corresponding PSF map. The spectral slit images are properly sized and tilted according to the sampling and tilt variation along each order, and scaled for the integrated spectral flux and efficiency. The code architecture has been developed in order to properly exploit the Matlab® Parallel toolbox functionalities, such that the different diffraction orders (or a fraction of them) can run in parallel. Once all the spectral images have been piled up to sum their photons distribution at sub-pixel scale, the synthetic frame is re-binned to pixel scale. Then photon noise and all other specific detector noises are added by the Detector Module.

### 3.8 Detector Module

The task of this module is to simulate, on the rendered data at pixel scale, the effects produced by the detector. These are QE, photon noise, read-out noise (RON), dark current (DC), bias levels, pixels non-uniform response (PRNU), pixel cross-talk (due to charge diffusion) and hot/bad pixels masks. In addition, this module also generates bias and dark frames for calibrations and add cosmic ray hits (see[1]). The QE curves for both the UV-VIS CCD and the NIR detector, see Fig. 15, have been extrapolated from both datasheet and measurements at specific wavelengths given by manufacturer; therefore these curves are representative of the detector performance but cannot be taken as fully real data.

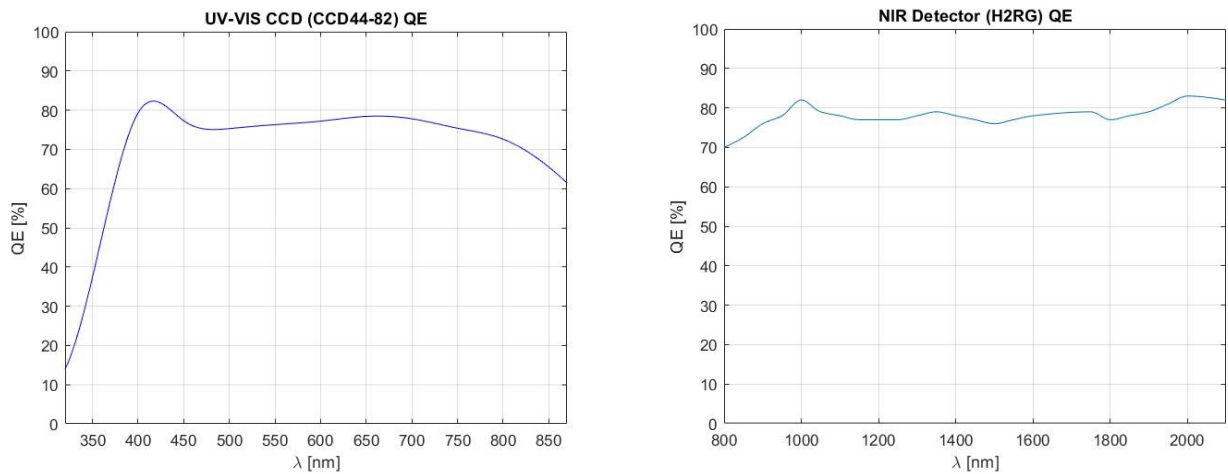

Figure 15. Spectrograph detectors QE. Left panel: UV-VIS CCD. Right panel NIR H2RG detector.

The parameters for some other relevant implemented effects are the followings; also these ones are extrapolated from datasheet, measurements by manufacturer tests or from same detector type in similar instrument.

|  | Gain (ADU/e-) | RON (rms e-) | DC (e/pix/s) | PNUR (%) | Cross-talk (%) |
|---|---|---|---|---|---|
| UV-VIS CCD | 1.25 | 2.3 | 2.7e-5 | 1.6 | 1 |
| NIR Detector | 0.5 | 11 (CDS reading) | < 1e-3 | 4 | 0.5 |

# 4. SIMULATED FRAMES

We present here some examples of simulated frames generated with this tool. A first example is the NIR synthetic image of a black body star at 5800 K with *V*-mag = 16 (Vega System). The observation has been simulated in new moon, at AM=1.2 with sky PWV = 10 mm. The seeing assumed is 0.9*arcsec*, at 550nm, and the selected slit width is 1*arcsec*. A single exposure with a detector integration time (DIT according to NIR nomenclature) of 600 sec is simulated. The different diffraction orders traces with the object centered in the slit and sky emission features are clearly visible, as well cosmic rays traces. In this simulation, the dark for the NIR detector is randomly generated following the histogram of dark pixels effectively measured on a similar real detector, re-scaled on the basis of a wide spatial scale variation across detector area. In this way, it is possible to mimic the spatial variation of warm/hot pixels in the dark frames usually seen in real instruments; this can be mostly noticed in the bottom part of the frame.

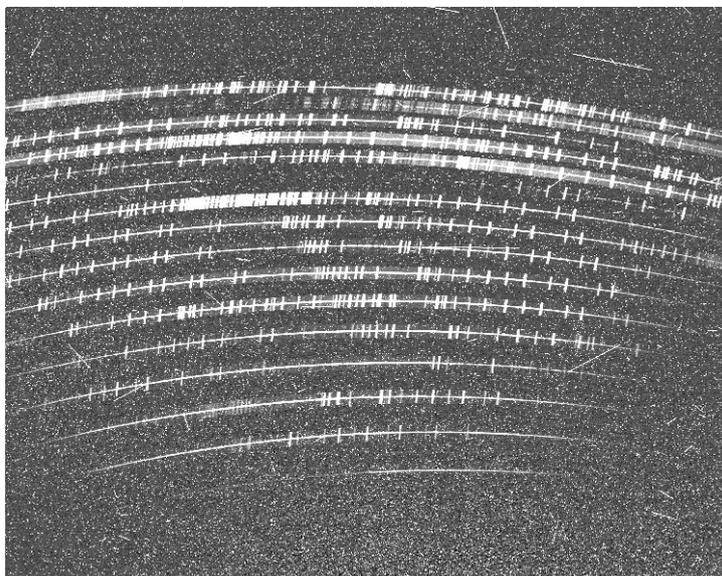

Figure 16. NIR arm synthetic frame of a science observation of black body star at 5800 K with *V*-mag = 16 (Vega System). See text for other simulation details.

Fig. 17 shows an example of a NIR wavelength calibration frame simulated with the multi pinhole mask and exposure of 10 sec. The spectral resolution elements related to the emission lines of the Pen-rays lamps along each order can be seen both in the full frame and in the zoomed box, where also the 9 pinholes images in cross dispersion direction can be distinguished. The wide spread pinhole images in dispersion direction, see the zoomed box, are blended emission lines.

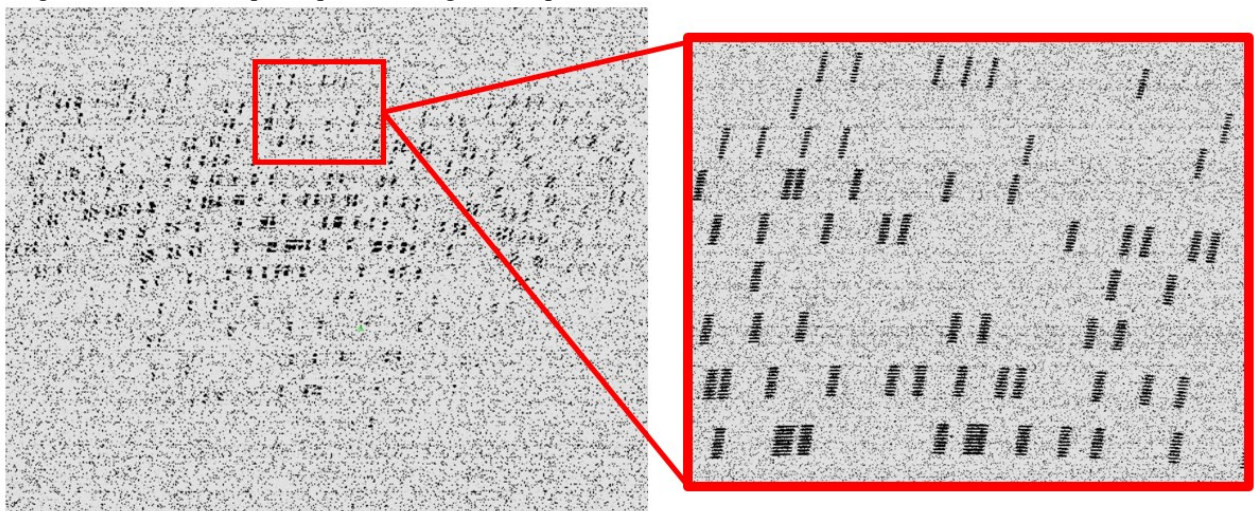

Figure 17. NIR wavelength calibration frames with PenRays (Ar-Hg-Ne-Xe) emission lines and multi-pinhole mask.

An example of how a UV-VIS science frame would looks like can be seen in Fig. 18. An observation of a G0V star, with *V*-mag = 18 (Vega System). The observation has been simulated again in new moon, at AM=1.2 with sky PWV = 10 mm. The seeing assumed is 0.87*arcsec*, at 550nm, and the selected slit width is 1*arcsec*. A single exposure of 1200 sec is emulated. The different quasi-orders traces with the object centered in the slit and sky emission features are clearly visible in the top and bottom quasi-orders, which are *r*-band and *i*-band respectively. While in the *u*-band (the second quasi-order from the top), the combined effect of star SED, instrument resolution, seeing and atmospheric transmission result in a trace that has 80% lower photon flux with respect to the *i*-band for example. The slit tilt of the different spectral resolution elements images can be visually noticed mostly in the bottom quasi-order (*i*-band) thanks to greater contrast and presence of sky emission lines. The presence of hot/bad pixels and cosmic rays, for this frame, has been uniformly simulated across detector area.

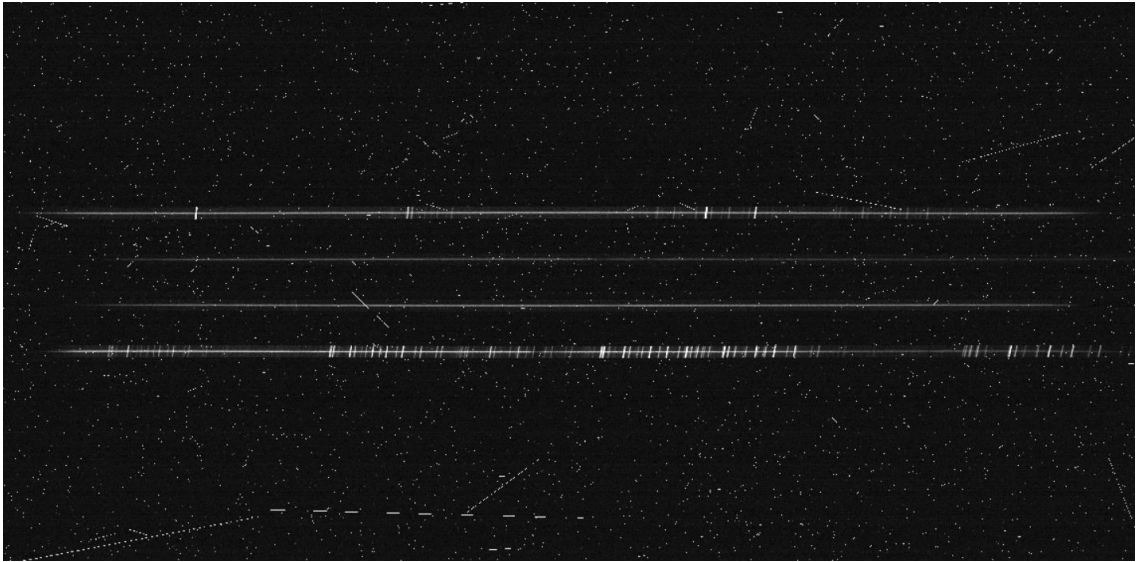

Figure 18. UV-VIS arm simulated synthetic frame of a science observation of a G0V star V-mag = 18 (Vega System). See text for other simulation details.

Fig. 19 shows a flat slit calibration frame in the UV-VIS, with QTH+D2 combined sources and 10 sec of exposure. The *r*-band and *g*-band (1st and 3rd from top respectively) have a sharp cut-off at the spectral range edge. The *u*-band exhibits a lower flux at the beginning of its wavelength range due to the CCD-QE (see Fig. 15, left panel), while the *i*-band is affected by the UV-VIS/NIR dichroic transition region above 800nm (see Fig. 8), as can be seen from the right side of its trace (the bottom one).

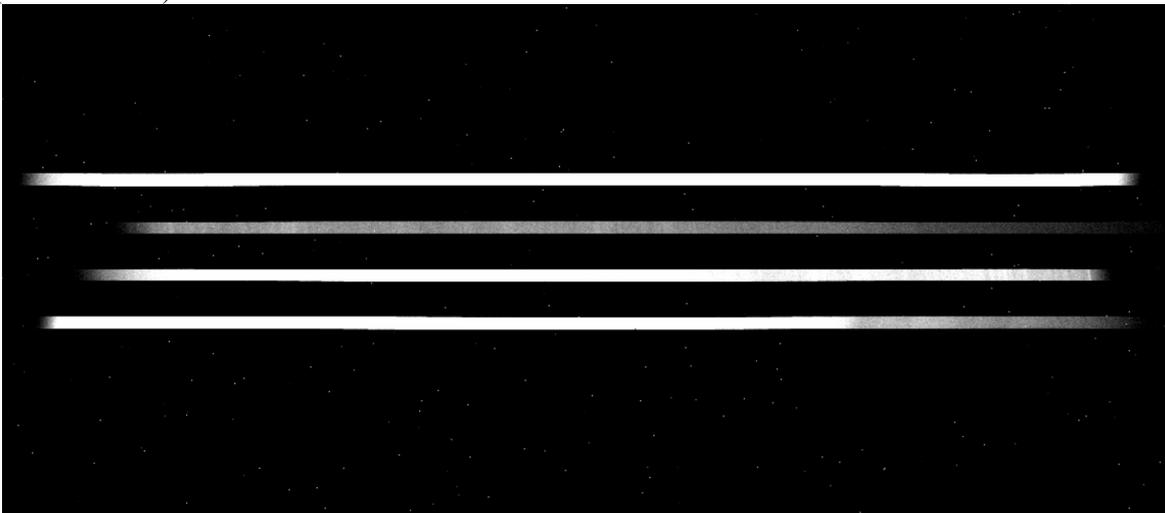

Figure 19. UV-VIS calibration frame. Flat slit with QTH+D2 lamps and 10 sec of exposure.

# 5. CONCLUSIONS

In this contribution we described the architecture and development of the NTT-SOXS E2E instrument simulator, showing both the different modules tasks and functionalities, and how the Image simulation Module produce a synthetic image for both the instrument arms.

Examples of simulations regarding both science observations and calibration frames have been presented. A specific set of calibration and observation synthetic raw frames will be generated and used both to aid (mainly in the UV-VIS arm) and cross-check (in the NIR arm) the DRS, which is currently under parallel development using real X-Shooter NIR arm frames (see for details[15]). The cross-check in the NIR arm will be fundamental to debug the simulator (if this will be required) and to validate a full end-to-end simulation, in which the ultimate goal is the production of synthetic 1D reduced spectra.